\documentclass[conference]{IEEEtran}
\IEEEoverridecommandlockouts

\usepackage{cite}
\usepackage{amsmath,amssymb,amsfonts}
\usepackage{algorithmic}
\usepackage{orcidlink}
\usepackage{graphicx}
\usepackage{textcomp}
\usepackage{enumitem} 
\usepackage{xcolor}
\usepackage{todonotes}
\usepackage{eqnarray}
\usepackage{stmaryrd}
\usepackage{lipsum}
\usepackage{nicematrix}
\usepackage{xcolor} 
\setuptodonotes{inline}
\usepackage[labelformat=simple]{subcaption}

\usepackage{pgf}
\usepackage{pgfplots}
\usepackage{tikz}
\usepackage{amsmath,amssymb}
\usepackage{stfloats}
\usepackage{tikz}
\usepackage{adjustbox}
\usetikzlibrary{
    arrows.meta,
    positioning,
    calc,
    fit,
    shadows
}
\usepackage[table]{xcolor}
\usepackage{colortbl}
\usepackage{comment}
\usepackage{ntheorem}

\usepackage{hyperref}
\usepackage{url}

\begin{document}

\title{Tensor-Based Modulation on the Unit Circle:\\ A Coding Perspective\\
\thanks{This work has been partially supported by the ANR under the “France 2030” program under grant NF-PERSEUS (ANR-22-PEFT-0004), and under grant WARM-M2M (ANR-24-CE25-2514).}
}

\author{\IEEEauthorblockN{Sweta Suresh}
\IEEEauthorblockA{
\textit{Inria, INSA Lyon,}\\
\textit{CITI Laboratory UR3720}\\
Villeurbanne, France}
\and 
\IEEEauthorblockN{Charly Poulliat}
\IEEEauthorblockA{\textit{University of Toulouse,} \\
\textit{INPT, IRIT Laboratory}\\
 Toulouse, France}
\and
\IEEEauthorblockN{Claire Goursaud}
\IEEEauthorblockA{
\textit{Inria, INSA Lyon,}\\
\textit{CITI Laboratory UR3720}\\
Villeurbanne, France}
\and
\IEEEauthorblockN{Maxime Guillaud}
\IEEEauthorblockA{
\textit{Inria, INSA Lyon,}\\
\textit{CITI Laboratory UR3720}\\
Villeurbanne, France}
}

\maketitle

\begin{abstract}
Tensor-based modulation (TBM) provides a multi-linear spreading framework for blind multi-user separation in unsourced random access. In this paper, we show that  TBM is a coded modulation built on a non-binary linear block code over $\mathbb{Z}_M$, whose symbols are mapped to 
$M$-PSK modulation, defining a geometrically uniform signal space code. We explicitly derive this generator matrix, characterize its rank deficiency, and show that reference symbols for tensor identifiability correspond to code shortening, producing a quasi-systematic or a systematic code, depending on the number of considered reference symbols for the TBM.
Simulations in single-user AWGN and multi-user non-coherent multi-antenna fading channels demonstrate strong robustness and interference resilience, establishing TBM as a scalable, algebraically structured modulation–coding scheme bridging tensor representations and modern coding theory.
\end{abstract}

\section{Introduction}
Unsourced random access (URA) \cite{Polyanskiy2017,liva2024unsourced} imposes unprecedented requirements on the decoder, requiring parallel interference cancellation (PIC) among a (typically) large number of users. Classical successive interference cancellation (SIC) do not perform well for URA, since inter-user interference arises from many weak users, hence no single interferer dominates, which prevents the capture effect which is at the core of SIC.

The tensor-based modulation (TBM) \cite{Decurninge2021} addresses this challenge through multi-linear spreading (MLS) scheme that couples information-bearing symbols through tensor products. Originally proposed for URA, TBM leverages the unique decomposability of low-rank tensors to enable blind multi-user separation. Its robustness to synchronization errors and Doppler effects has been established in \cite{TBM_timing_offset_Globecom22,wcnc2026_TBM_Doppler_estimation}.

Recent work focused on the encoding of phase shift keying (PSK) symbols with TBM, is denoted as TBM-PSK. 
An efficient belief-propagation (BP) decoder for TBM-PSK over non-coherent fading channels was introduced in \cite{Suresh_IZS2026_Riemannian_BP}.
It relies on relaxing discrete $M$-PSK symbols to continuous unit-modulus variables on the complex circle, with messages parameterized by von Mises distributions (vM-BP). 
This formulation allows approximate closed-form message updates and has the appealing property that both equalization and decoding naturally operate on the same unit-circle manifold, avoiding enumerating over constellation points during bit log-likelihood/complex symbols conversion.

In this work, we show that  TBM is a coded modulation built on a non-binary linear block code defined over $\mathbb{Z}_M$ whose symbols are mapped to 
$M$-PSK modulation, defining a geometrically uniform signal space code \cite{LoeligerPhD,LoeligerTIT91,SidharaPhD}. The code is explicitly defined through its generator matrix. Interpreting it as a kind of non-binary "low density" generator matrix (LDGM) code \cite{richardson2008modern} over $\mathbb{Z}_M$ enables decoding of TBM using message passing algorithms, operating on the corresponding Tanner/factor graph. Although TBM is fundamentally tailored for multi-user communications, we present a single-user analysis that mainly serves to enlighten and formalise the algebraic structure imposed by TBM. 

The contributions are:
\begin{itemize}
\item We show that when $M$-PSK information symbols are encoded using the MLS defining TBM, the resulting sequences can be interpreted as codewords generated by a linear block code over $\mathbb{Z}_M = \{0,1, \ldots, M-1\}$ under addition modulo $M$, followed by an $M$-PSK modulation.
\item  We characterize the rank deficiency of the corresponding generator matrix, and show that reference symbols required for tensor identifiability correspond to code shortening, producing a quasi-systematic or a systematic code, depending on the number of considered reference symbols for the TBM. 
\item We demonstrate that TBM-PSK combined with the vM-BP approach can form the basis of efficient parallel interference cancellation decoders.

\end{itemize}

\section{Background on Tensor-Based Modulation} \label{Code_construcation_background_on_tbm}
To clarify the code construction, we first introduce the MLS approach from \cite{Decurninge2021} in its original context of URA over the multiple access channel. At transmitter $k$, a $T$ length symbol vector $\mathbf{s}^{(k)}\in\mathbb{C}^T$ is generated as 
\begin{equation} \label{MLspreading}
\mathbf{s}^{(k)}=\mathbf{x}_1^{(k)} \otimes \mathbf{x}_2^{(k)} \otimes \cdots \otimes \mathbf{x}_d^{(k)} 
\end{equation}
where $\mathbf{x}_i^{(k)} \in \mathbb{C}^{T_i}, i=1\ldots d,$ are information bearing vectors, $\prod_i T_i = T$ is one possible factorization of the blocklength $T$ with $T_i\in \mathbb{N}\backslash \{1\}$, and $\otimes$ denotes the Kronecker product.

In the URA case, we consider a single-input multiple-output (SIMO) multiple-access fading channel with $K_a$ users and $N_r$ receive antennas
\begin{equation}
    \mathbf{y} = \sum_{k=1}^{K_a} \mathbf{s}^{(k)} \otimes \mathbf{h}^{(k)} + \mathbf{n},
    \label{TBM-s-h}
\end{equation}
with $\mathbf{y}, \mathbf{n} \in \mathbb{C}^{N_r T}$, and where $\mathbf{h}^{(k)} \in \mathbb{C}^{N_r}$ denotes the state of the channel of user $k$, and $\mathbf{n}$ is an additive white Gaussian noise (AWGN) term.
In this context, the noise-free received signal from $K_a$ active users is given by
\begin{equation}\label{low_rank_tensor}
    \mathbf{y}_0 = \sum_{k=1}^{K_a} \mathbf{x}_1^{(k)} \otimes \mathbf{x}_2^{(k)} \otimes \cdots \otimes \mathbf{x}_d^{(k)} \otimes \mathbf{h}^{(k)},
\end{equation}
which corresponds to a $d+1$-order tensor of size $T_1 \times \cdots \times T_d \times N_r$, each rank one term representing one user. 
Previous works \cite{Decurninge2021,fang2025polarTBM_soft_decoding} have shown that tensor algebra \cite{kolda2009tensor} enables blind multi-user separation and channel estimation through the canonical polyadic (CP) decomposition of low-rank tensors. The uniqueness properties of CP decomposition \cite{chiantini14} further characterize the high-SNR behaviour and determine the maximum number of users that can be reliably resolved.
For each rank-1 component in \eqref{low_rank_tensor}, unicity is defined up to permutation and complex scalar ambiguities. Consequently, each vector $\mathbf{x}_i^{(k)}$ embeds one reference symbol and $T_i-1$ information symbols. 
Without loss of generality, the reference symbol is fixed as $x_{i, 1}^{(k)}=1$, where $x_{i, j}^{(k)}$ denotes the $j$-th scalar element of $\mathbf{x}_i^{(k)}$.
This configuration with one pilot per mode is required in the context of non-coherent multi-user random access, as in \cite{Decurninge2021}.
In this work, we will consider the following cases:
\begin{description}
\item[Case 1:] one reference symbol per tensor mode;
\item[Case 2:] no reference symbols;
\item[Case 3:] $d-1$ reference symbols. 
\end{description}

\section{TBM as a Coded Modulation}\label{coding_perspective}
In this section, we show that the MLS \eqref{MLspreading} can be equivalently described as a linear map over $\mathbb{Z}_M$.
Leveraging this perspective, TBM can be interpreted as a structured non-binary coded modulation scheme. 
\subsection{Preliminaries}\label{preliminaries}
Building on the configuration described above with \textit{Case 2}, we re-interpret the MLS operation \eqref{MLspreading}, in the special case where the information symbols belong to a PSK modulation of order $M$,
$\mathcal{X}_M=\left\{\left.e^{j\left(\frac{2 \pi v}{M}\right)} \right\rvert\, v=0, \ldots, M-1\right\} \subset \mathbb{C}$.
Let $\mathbb{Z}_M = \{0,1, \ldots, M-1\}$ denote the additive cyclic group under modulo-$M$ addition, isomorphic to the commutative group $\mathbb{Z} / M \mathbb{Z}$, i.e. $\left(\mathbb{Z}_M,+\bmod M\right) \cong(\mathbb{Z} / M \mathbb{Z},+)$. The mapping 
\begin{equation}
    \begin{split}
        \varphi: &   \;\mathbb{Z}_M \longrightarrow \mathcal{X}_M\\
        & v  \longmapsto \; \varphi(v)= e^{j\left(\frac{2 \pi v}{M}\right)} \\
    \end{split}
\end{equation}
defines an isomorphism from  $\left(\mathbb{Z}_M,+\bmod M\right)$ to $(\mathcal{X}_M,\times )$.

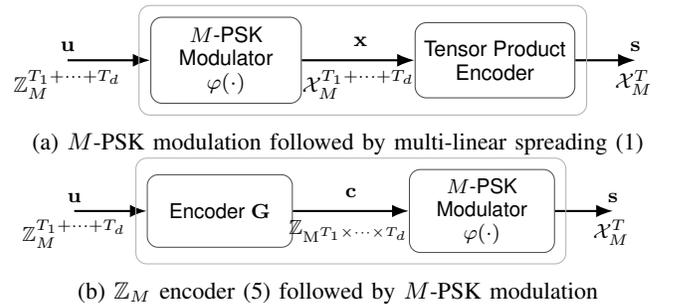
\begin{figure}[b]
  \begin{subfigure}{\columnwidth}
  \adjustbox{trim=0 0.03mm 0 0,clip}{\begin{tikzpicture}[
    font=\sffamily\footnotesize,
    node distance=15mm,
    every node/.style={transform shape}
]

\tikzset{	
    block/.style={
        draw=black!70,
        rounded corners=1.5mm,
        fill=white,
        minimum height=10mm,
        minimum width=20mm,
        align=center
    },
    conn/.style={-Latex, thick},
    lab/.style={font=\sffamily\footnotesize}
}


\node[block] (mod) {$M$-PSK\\Modulator\\$\varphi(\cdot)$};
\node[block, right=of mod] (enc) {Tensor Product\\Encoder};

\coordinate (in) at ($(mod.west)+(-11mm,0)$);
\draw[conn] (in) -- (mod.west);

\node[lab, above] at (in)
{$\mathbf{u}$};
\node[lab,below] at (in) {$\mathbb{Z}_M^{T_1+\cdots+T_d}$};
\draw[conn] (mod) -- (enc);
\node[lab, above=2pt, align=center] at ($(mod)!0.5!(enc)$)
{$\mathbf{x}$};
\node[lab,below] at ($(mod)!0.5!(enc)$) {$\!\mathcal{X}_M^{T_1+\cdots+T_d}$};
\coordinate (out) at ($(enc.east)+(8mm,0)$);
\draw[conn] (enc.east) -- (out);

\node[lab, above] at (out)
{$\mathbf{s}$};
\node[lab,below] at (out)
{$\!\mathcal{X}_{M}^{T}$};

\node[
    draw=gray!60,
    rounded corners=1.5mm,
    fit=(mod)(enc),
    inner sep=1.5mm
] {};


\end{tikzpicture}}
  \caption{$M$-PSK modulation followed by multi-linear spreading \eqref{MLspreading}}
  \label{PSK-then-MLS}
  \end{subfigure}
  \begin{subfigure}{\columnwidth}
  \adjustbox{trim=0 0.03mm 0 0,clip}{\resizebox{0.95\textwidth}{!}{
\begin{tikzpicture}[
    font=\sffamily\footnotesize,
    node distance=16mm,
    every node/.style={transform shape}
]

\tikzset{	
    block/.style={
        draw=black!70,
        rounded corners=1.5mm,
        fill=white,
        minimum height=10mm,
        minimum width=20mm,
        align=center
    },
    conn/.style={-Latex, thick},
    lab/.style={font=\sffamily\footnotesize}
}


\node[block] (mod) {Encoder $\mathbf{G}$};
\node[block, right=of mod] (enc) {$M$-PSK\\Modulator\\$\varphi(\cdot)$};

\coordinate (in) at ($(mod.west)+(-10mm,0)$);
\draw[conn] (in) -- (mod.west);

\node[lab, above] at (in)
{$\mathbf{u}$};
\node[lab,below] at (in) {$\mathbb{Z}_M^{T_1+\cdots+T_d}$};
\draw[conn] (mod) -- (enc);
\node[lab, above=2pt, align=center] at ($(mod)!0.5!(enc)$)
{$\mathbf{c}$};
\node[lab,below] at ($(mod)!0.5!(enc)$) {$\!\mathbb{Z}_{\mathrm{M}^{T_1\times \cdots \times T_d}}$};
\coordinate (out) at ($(enc.east)+(8mm,0)$);
\draw[conn] (enc.east) -- (out);

\node[lab, above] at (out)
{$\mathbf{s}$};
\node[lab,below] at (out)
{$\!\mathcal{X}_{M}^{T}$};

\node[
    draw=gray!60,
    rounded corners=1.5mm,
    fit=(mod)(enc),
    inner sep=1.5mm
] {};


\end{tikzpicture}}}
  \caption{$\mathbb{Z}_M$ encoder \eqref{eq_GMencoding} followed by $M$-PSK modulation} 
  \label{encoderG-then-PSKmod}
  \end{subfigure} 
  \caption{Two equivalent interpretations of PSK-TBM: MLS, and coded modulation.} \label{fig_equivalent_models}
\end{figure}

For brevity, in the sequel we drop the superscript $(k)$ since we focus on a single user. 
Let $\mathbf{u}_i \in \mathbb{Z}_M^{T_i}$ denote the integer valued vector of $T_i$ symbols encoded in $\mathbf{x}_i$, for all $i\in \llbracket 1, d \rrbracket$. The classical TBM encoding process consists in a PSK modulation followed by MLS as per \eqref{MLspreading}, i.e. $\mathbf{x}_i = \varphi\left( \mathbf{u}_i\right)$ (applying $\varphi$ element-wise) followed by 
 $\mathbf{s}=\varphi\left( \mathbf{u}_1\right) \otimes \cdots \otimes \varphi\left( \mathbf{u}_d\right)$. This process is depicted in Fig.~\ref{PSK-then-MLS}. 

The transmitted symbols can equivalently be obtained as $\mathbf{s}=\varphi\left(\mathbf{u}\mathbf{G} \right)$ for some encoding matrix $\mathbf{G} \in \mathbb{Z}_M^{K \times T}$,  where $\mathbf{u} = \left[ \mathbf{u}_1, \cdots, \mathbf{u}_d  \right]$ contains the  $K \triangleq \sum_i T_i$ symbols transmitted.
This indicates that the encoding process can equivalently be interpreted as a non-binary encoding on $\mathbb{Z}_M$ followed by $M$-PSK modulation, as depicted in Fig.~\ref{encoderG-then-PSKmod}. 
A practical consequence is that the resulting code is geometrically uniform
\cite{Forney_geometrically_uniform_codes_TIT91}; geometric uniformity implies that the Voronoi regions associated to all codewords have the same shape.
We now establish the structure of the encoding matrix. 

\subsection{Equivalent (non minimal) $\mathbb{Z}_M$ Encoder}
For \textit{Case 2}, $\mathbf{u}$ contains a maximum of $\sum_i T_i$ information-bearing symbols on $\mathbb{Z}_M$. By doing so, we do not put an a priori on the structure of the information-bearing symbols. Under this configuration, we will see in Section \ref{sec:CSI} that the resulting encoding matrix $\mathbf{G}$ is a non-minimal generator matrix, i.e. $\mathrm{rank}( \mathbf{G})=K' < K$, implying that $\mathbf{u}$ is not uniquely identifiable from a codeword $\mathbf{c}$.
The encoding operation in Fig. \ref{PSK-then-MLS}, can be rewritten as  a non-systematic linear encoding operation using a generator matrix $\mathbf{G}$ of size $K \times T$, i.e.
\begin{equation}\label{eq_GMencoding}
    \mathbf{c}=[c_1,c_2,\cdots, c_T]=\mathbf{u}\mathbf{G}.
\end{equation}
Here, $\mathbf{G}$ follows the same structure of Gallager's modulo-$M$ (non-binary) LDPC codes \cite{Gallagerthesis}, having all non-zero positions equal to $1$. 
Since $\mathbf{s}$ is the vectorized form of a tensor of dimension $T_1 \times \cdots \times T_d$, it is possible to index its $T$ elements using a $d$-tuple $(m_1, m_2, \cdots, m_d) \in \llbracket 1, T_1 \rrbracket \times \cdots \times \llbracket 1, T_d \rrbracket$ representing the $d$-dimensional index of a tensor element \cite[Sec.~2.4]{kolda2009tensor}.
Using e.g. the lexicographic ordering, we can bijectively map each $p\in \llbracket 1, T \rrbracket$ to a tuple $(m_1, m_2, \cdots, m_d)$; which we denote by $p \leftrightarrow (m_1, m_2, \cdots, m_d)$.
According to \eqref{MLspreading}, the $p$-th element of $\mathbf{s}$ can be written as
\begin{equation}
\begin{split}
    s_{p} & \leftrightarrow  s_{m_1, m_2, \cdots, m_d}\\
    & =x_{1,m_1}x_{2,m_2} \cdots x_{d,m_d}, \forall 1 \leq m_i \leq T_i.\\
\end{split} 
\end{equation}

Note that the indices are related to the positions within each vector component that participate in the $\mathbb{Z}_m$ check-sum of the symbol word position $p$.  Let 
\begin{equation}
    \begin{matrix}
\mathbf{e}_{m_i}^i &= 
        [0,\ldots,0&, 1,& 0,\ldots,0],\\
        &&\uparrow&  \\
        &&m_i&      & \\
\end{matrix}
\end{equation}
be the binary indicator vector of length $T_i$ associated with the position index $m_i$ related to the vector component $\mathbf{u}_i$. Then for each $p \in \llbracket 1 ,T \rrbracket$, the $d$-tuple $(m_1,\ldots, m_d)$ can be expanded into  its binary expansion given by $(\mathbf{e}_{m_1}^1, \ldots,\mathbf{e}_{m_d}^d)$. Therefore, for each $p$, or equivalently $(m_1,\cdots, m_d)$, we have
\begin{equation}
    \begin{split}
  & c_p \triangleq c_{m_1, m_2, \cdots, m_d}\\ 
  & = u_{1, m_1} + u_{2, m_2}+ \cdots+ u_{d, m_d} \bmod M \\
  & =\mathbf{u}_1(\mathbf{e}_{m_1}^{1})^{\top}+\mathbf{u}_2(\mathbf{e}_{m_2}^{2})^{\top}+ \cdots+\mathbf{u}_d(\mathbf{e}_{m_d}^d)^{\top}\bmod M  \\
  &=\mathbf{u}  \mathbf{g}_p^{\top}\bmod M ,
\end{split}
\end{equation}
where $\mathbf{g}_p = [\mathbf{e}_{m_1}^1, \cdots, \mathbf{e}_{m_d}^2]$.
Hence each codeword symbol $c_p$ is the modulo-$M$ sum of input symbols from the concatenated vector $\mathbf{u}$ for which we have a corresponding one in the $p$-th row of $\mathbf{G}$. This generator matrix is a particular case of a code over $\mathbb{Z}_M$ as all non-zero elements are the unity element $1$. 
Thus, $\mathbf{c} \triangleq [c_1, c_2, \cdots, c_{T}]$ can be written as
\begin{equation}
\begin{split}
    \mathbf{c}=\mathbf{u} \begin{bmatrix}
                    \mathbf{g}_1^{\top} &\mid \mathbf{g}_2^{\top} &\mid \cdots &\mid \mathbf{g}_T^{\top}
                    \end{bmatrix} =  \mathbf{u} \mathbf{G}.
\end{split}
\end{equation}
Hence, it can be concluded that $\mathbf{G}=\begin{bmatrix}
                    \mathbf{g}_1^{\top} &\mid \mathbf{g}_2^{\top} &\mid \cdots &\mid \mathbf{g}_T^{\top}
                    \end{bmatrix}.$\\
                    
As an example, for $\left(T_1, T_2, T_3\right) =(4,2,2)$, we have
$\forall p \in \llbracket 1,16 \rrbracket, \; p=1+4\left(m_1-1\right)+2\left(m_2-1\right)+1\left(m_3-1\right)$ and thus, the matrix $\mathbf{G}$ is given by
\begin{equation}
\scalebox{0.65}{$
    \mathbf{G}=
    \left[
\begin{array}{cccccccccccccccc}
1&1&1&1&0&0&0&0&0&0&0&0&0&0&0&0\\
0&0&0&0&1&1&1&1&0&0&0&0&0&0&0&0\\
0&0&0&0&0&0&0&0&1&1&1&1&0&0&0&0\\
0&0&0&0&0&0&0&0&0&0&0&0&1&1&1&1\\
1&1&0&0&1&1&0&0&1&1&0&0&1&1&0&0\\
0&0&1&1&0&0&1&1&0&0&1&1&0&0&1&1\\
1&0&1&0&1&0&1&0&1&0&1&0&1&0&1&0\\
0&1&0&1&0&1&0&1&0&1&0&1&0&1&0&1
\end{array}
\right].\\
$}
\label{first_G_matrix}
\end{equation}

This matrix is a structured non-systematic generator defined over $\mathbb{Z}_M$.
Further, it can be shown that the matrix $\mathbf{G}$ can also be constructed using a recursive Kronecker construction. Let $\mathcal{T}^d=[T_1,T_2, \cdots, T_d]$ be the ordered $d$-tuples, such that $T_1 \geq T_2 \geq  \cdots\geq  T_d$, and $\mathcal{T}^n=[T_{d-n+1}, \cdots, T_d]$. Then, $\mathbf{G}_{\mathcal{T}^1}=\mathbf{I}_{T_d}$,
and $ \forall n \in \llbracket2,d \rrbracket$,
\begin{equation}
    \large \mathbf{G}_{\mathcal{T}^n}=\begin{bmatrix}
    \mathbf{I}_{t_n} \otimes \mathbf{1}_{t'_{n-1}} \\
    \mathbf{1}_{t_n} \otimes \mathbf{G}_{\mathcal{T}^{n-1}}
\end{bmatrix},
\end{equation}
where $  \mathbf{1}_{t}=[1, \cdots, 1] \in \mathbb{Z}_M^t$ and $t_n=T_{d-n+1},t'_n=\prod_{k=d-n+2}^{d}{T_k} $. Finally, $\mathbf{G}=\mathbf{G}_{\mathcal{T}^d}.$

\section{Information dimension identification: The Need for Reference Symbols} \label{need_for_reference}
\label{sec:CSI}

In Section \ref{coding_perspective}, we derived the equivalent encoding of TBM over $\mathbb{Z}_M$ for \textit{Case 2}, i.e. without analyzing the proper code space and information dimension. However, it is important to identify the true information code dimension of the underlying coded scheme and the proper subspace of $\mathbb{Z}_M^T$ defining the underlying code $\mathcal{C}$ defined over $\mathbb{Z}_M$.
To do so, the algebraic properties of the generator matrix $\mathbf{G}$ are studied next.

\subsection{Reference symbols in the coherent setting}
According to tensor properties in the analog domain, we know that the matrix $\mathbf{G}$ should be rank-deficient (the information code dimension should be less due to inherent constraints of the TBM structure) \cite{Decurninge2021}. By analyzing the structure of the generator matrix and considering a tensor defined by an ordered $d$-tuple, we can show that we can find exactly $d-1$ redundant rows that can be 'cancelled'. 
This can be shown by remarking that
the sum of the rows in $\mathbf{G}$ associated with a given tensor mode $i$ is the all-ones vector, i.e.
 \begin{equation}
     \forall i \in \llbracket 1, d \rrbracket, \sum_{l=T_{i-1}+1}^{T_i} \mathbf{r}_l= \left[1, \dots 1 \right], \; T_0=0,
 \end{equation}
where $\mathbf{r}_l$ is the $l$-th row of $\mathbf{G}.$
Thus, for the $d-1$ rows with row index $k \in \{(\sum_{m=1}^{i} T_m)+1, i \in \llbracket 1, d-1\rrbracket \}$, if we update row $k$ corresponding to the $i$ component vector as
\begin{equation}\label{eq_elimination}
    \mathbf{r}_k \leftarrow \sum_{m'=k}^{k+T_i-1}\mathbf{r}_{m'}  -\sum_{m=1}^{T_1}\mathbf{r}_m \; \mod M,
\end{equation}
then $ \mathbf{r}_k =\left[0,\cdots,0 \right].$ 
Note that the choice of $k$ inside a tensor mode is arbitrary; any other choice of one row per mode would lead to an equivalent form.
Here we follow the convention from \cite{Decurninge2021} and choose the first position in a mode for reference symbols.
It can be shown that up to $d-1$ rows from $\mathbf{G}$ can be eliminated as in \eqref{eq_elimination}.
Thus we select the first rows in $d-1$ out of $d$ modes to serve as \emph{reference symbols}, hence the corresponding symbols cannot be used to bear information, and are fixed to $0_{\mathbb{Z}_M}$ (equivalently to $1$ in the constellation domain). 
This configuration -- \textit{Case 3} with $d-1$ reference symbols, in each but one mode -- is sufficient for the coherent (single or multi-user) channel setting. 
Let $\mathbf{u}_r=\left[\mathbf{u}_1, \mathbf{u}^r_2 \cdots, \mathbf{u}^r_d\right] \in \mathbb{Z}_M^{K_r}$ be the reduced information vector, containing $K_r=\sum_i{T_i}-d+1$ symbols, obtained by removing the fixed reference symbols.
The code associated with the TBM in \textit{Case 3} is given by 
\begin{equation}
    \mathcal{C}_{\mathrm{TBM}}^{\mathrm{coh}}=\left\{\mathbf{c} \in \mathbb{Z}_M^{T} \mid \mathbf{c}=  \mathbf{u}_r \mathbf{G}_r; \mathbf{u}_r \in \mathbb{Z}_M^{K_r}\right \},
\end{equation}
where $\mathbf{G}_r$ of size $K_r \times T$, is the reduced generator matrix obtained from $\mathbf{G}$ by removing the rows corresponding to reference symbols set to zero. It is easy to see that the defined code is partially systematic, as we have exactly $T_1$ systematic positions associated to the symbols of the first mode. The corresponding rate is $\eta^{\mathrm{coh}}=\frac{K_r}{T} \log _2(M).$
For example \eqref{first_G_matrix} with $\left(T_1, T_2, T_3\right) =(4,2,2)$, we have
\begin{equation}
    \scalebox{0.70}{$
    \mathbf{G}_r=
    \left[
\begin{array}{cccccccccccccccc}
1 & 1 & 1 & 1 & 0 & 0 & 0 & 0 & 0 & 0 & 0 & 0 & 0 & 0 & 0 & 0 \\
0 & 0 & 0 & 0 & 1 & 1 & 1 & 1 & 0 & 0 & 0 & 0 & 0 & 0 & 0 & 0 \\
0 & 0 & 0 & 0 & 0 & 0 & 0 & 0 & 1 & 1 & 1 & 1 & 0 & 0 & 0 & 0 \\
0 & 0 & 0 & 0 & 0 & 0 & 0 & 0 & 0 & 0 & 0 & 0 & 1 & 1 & 1 & 1 \\
0 & 0 & 1 & 1 & 0 & 0 & 1 & 1 & 0 & 0 & 1 & 1 & 0 & 0 & 1 & 1 \\
0 & 1 & 0 & 1 & 0 & 1 & 0 & 1 & 0 & 1 & 0 & 1 & 0 & 1 & 0 & 1
\end{array}
\right].$}
\end{equation}

In addition, we can also define the so-called generalized parity-check matrix \cite{Yedidia2002GeneratingCR} given by $\tilde{\mathbf{H}}=\left[\mathbf{G}_r^{\top} \; \mid \; (M-1)\mathbf{I}_{T}\right]$,
for which we can define $\tilde{\mathbf{c}}=[\mathbf{u}_r,\mathbf{c}],$ such that $\tilde{\mathbf{c}}\tilde{\mathbf{H}}^{\top}=\mathbf{0}.$

\subsection{Reference symbols in the non-coherent setting}
For the \emph{non-coherent} block-fading channel scenario, an extra reference symbol is required to resolve the scalar ambiguity due to the channel fading coefficient, as discussed in Sec.~\ref{Code_construcation_background_on_tbm}, leading to \textit{Case 1}.
Fixing the reference symbols to $0 \in \mathbb{Z}_M$ in the code domain can be interpreted as code-symbols used for shortening.
This is equivalent to removing the symbols corresponding to known zero-valued symbols from $\mathbf{u}$ or $\mathbf{u}_r$.
Each remaining symbol position then gains a systematic observation that aids decoding. 
$c_1$, which represents the product of $d$ reference symbols, can also be treated separately since it carries no information.

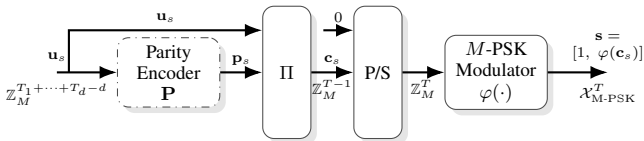
\begin{figure}[h]
\centering
\begin{tikzpicture}[
    scale=0.80,
    transform shape,
    node distance=7mm,
    every node/.style={transform shape}
]

\tikzset{	
    block/.style={
        draw=black!70,
        rounded corners=1.5mm,
        fill=white,
        minimum height=10.5mm,
        minimum width=17mm,
        align=center,
        font=\small,
        drop shadow={shadow xshift=0.8mm, shadow yshift=-0.8mm, opacity=.2}
    },
    bigblock/.style={
        draw=black!70,
        rounded corners=1.5mm,
        fill=white,
        minimum height=22mm,
        minimum width=8mm,
        align=center,
        font=\small,
        drop shadow={shadow xshift=0.8mm, shadow yshift=-0.8mm, opacity=.2}
    },
    conn/.style={-Latex, thick},
    lab/.style={font=\scriptsize}
}

\node[block, dash dot] (dem) {Parity\\ Encoder\\ $\mathbf{P}$};
\node[bigblock, right=of dem] (INT) {$\Pi$};
\node[bigblock, right=of INT] (MUX) {P/S};
\node[block, right=of MUX] (dec) {$M$-PSK \\Modulator\\ $\varphi(\cdot)$};

\draw[conn] (dem) -- node[lab, above] {$\mathbf{p}_s$} (INT);

\draw[conn] (INT) -- 
node[lab, above] {$\mathbf{c}_s$}
node[lab, below] {$\mathbb{Z}_M^{T-1}$} (MUX);

\draw[conn] (MUX) -- 
node[lab, below] {$\mathbb{Z}_M^{T}$} (dec);

\draw[conn]
($(INT.east)+(2mm,7mm)$) --
node[lab, above] {$0$}
($(MUX.west)+(0mm,7mm)$);

\coordinate (in)  at ($(dem.west)+(-10mm,0)$);
\coordinate (out) at ($(dec.east)+(10mm,0)$);
\coordinate (in2)  at ($(dem.west)+(-8mm,0)$);

\draw[conn] (in) -- (dem.west);
\draw[conn] (dec.east) -- (out);
\draw[conn]
(in2) -- ($(in2)+(0,7mm)$) --
node[lab, above] {$\mathbf{u}_s$}
($(INT.west)+(0mm,7mm)$);

\node[lab, above=2pt] at (in)
{$\mathbf{u}_s$};

\node[lab, below=2pt] at (in)
{$\mathbb{Z}_M^{T_1+\cdots+T_d-d}$};

\node[lab, above=2pt] at (out)
{\shortstack{$\mathbf{s}=$\\
$[1,\;\varphi(\mathbf{c}_s)]$}};

\node[lab, below=2pt] at (out)
{$\mathcal{X}_{\mathrm{M\text{-}PSK}}^{T}$};

\end{tikzpicture}
\caption{Equivalent systematic encoding of TBM-PSK.}
\label{final_tbm}
\end{figure}

In matrix form, the additional reference symbol required for the non-coherent case 
consists in setting the first row of $\mathbf{G}_r$ to zero, leading to a rank-deficient matrix, as both the first row and column becomes zero. Removing first row and column, we get a generator matrix $\mathbf{G}_s$  of size $K_s \times (T-1)$, with $K_s=\sum_i\left(T_i-1\right).$ $\mathbf{G}_s$  is used to map the information vector $\mathbf{u}_s=\left[\mathbf{u}_1^r, \mathbf{u}_2^r \cdots, \mathbf{u}_d^r\right] \in \mathbb{Z}_M^{K_s}$ to a vector $\mathbf{c}_s \in \mathbb{Z}_M^{(T-1)},$ corresponding to all coded TBM symbols except $c_1$ which is fixed to $0$. 
TBM-PSK in \textit{Case 1} is thus equivalent to the code
\begin{equation}\label{eq_code_non_coherent}
\mathcal{C}_{\mathrm{TBM}}^{\mathrm{ncoh}}=\left\{\mathbf{c} \in \mathbb{Z}_M^T \mid \mathbf{c}=[0,\mathbf{u}_s \mathbf{G}_s]; \mathbf{u}_s \in \mathbb{Z}_M^{K_s}\right\},
\end{equation}
yielding a rate $\eta^{\mathrm{ncoh}} = \frac{K_s}{T} \log_2 M$ bits/channel use. $\mathbf{G}_s$ can be shown to be systematic with positions that are easily identified according to the positions of reference symbols.
For the example from \eqref{first_G_matrix} with $\left(T_1, T_2, T_3\right) =(4,2,2)$, we have
\begin{equation}
    \scalebox{0.7}{$
    \mathbf{G_s}=
    \left[
\begin{array}{cccccccccccccccc}
\textcolor{blue}{0}  & \textcolor{blue}{0}  & 0 & \textcolor{blue}{1} & 1 & 1 & 1 & \textcolor{blue}{0}  & 0 & 0 & 0 & \textcolor{blue}{0}  & 0 & 0 & 0 \\
 \textcolor{blue}{0}  & \textcolor{blue}{0}  & 0 & \textcolor{blue}{0}  & 0 & 0 & 0 & \textcolor{blue}{1} & 1 & 1 & 1 & \textcolor{blue}{0}  & 0 & 0 & 0 \\
 \textcolor{blue}{0}  & \textcolor{blue}{0}  & 0 & \textcolor{blue}{0}  & 0 & 0 & 0 &\textcolor{blue}{0}  & 0 & 0 & 0 & \textcolor{blue}{1} & 1 & 1 & 1 \\
 \textcolor{blue}{0} & \textcolor{blue}{1} & 1 & \textcolor{blue}{0}  & 0 & 1 & 1 & \textcolor{blue}{0} & 0 & 1 & 1 & \textcolor{blue}{0}  & 0 & 1 & 1 \\
\textcolor{blue}{ 1} &\textcolor{blue}{0}  & 1 & \textcolor{blue}{0}  & 1 & 0 & 1 & \textcolor{blue}{0}  & 1 & 0 & 1 & \textcolor{blue}{0}  & 1 & 0 & 1
\end{array}
\right].
$}
\label{G_s}
\end{equation}
Columns 1, 2, 4, 8 and 12 have a single non-zero element, and correspond to $K_s$ systematic positions after shortening. 
The reference symbols enable systematic encoding of the remaining information symbols. Consequently, we can derive an efficient systematic encoding scheme for \textit{Case 1}, where the generator matrix in systematic form is denoted $\mathbf{G}_{\mathrm{sys}}=[\mathbf{I}_{K}\mid\mathbf{P}]$, after appropriate column permutation $\Pi$. The corresponding encoding process is depicted in Fig.~\ref{final_tbm}.

\section{Factor Graph Representation and message passing decoding}

\begin{figure}[h]
  \centering
  \includegraphics[width=0.49\textwidth]{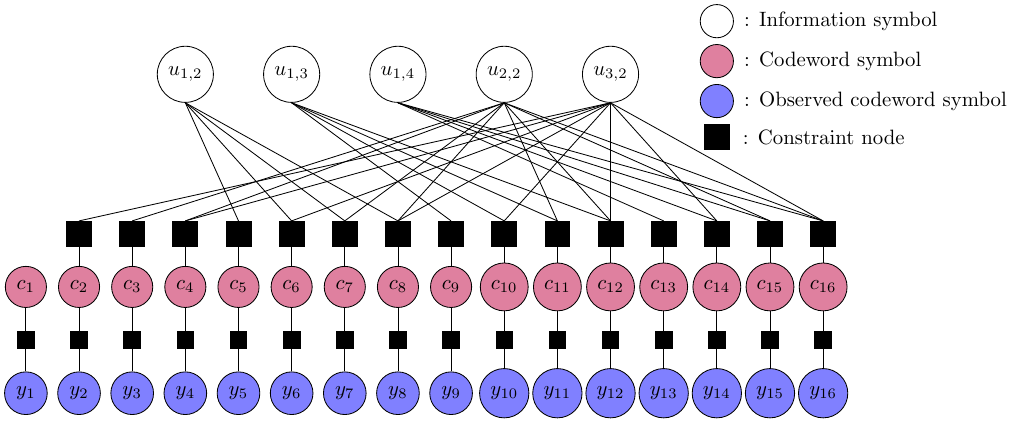}
  \caption{Factor graph representation of the proposed code construction for $\left(T_1, T_2, T_3\right) =(4,2,2)$, \textit{Case 1}. Note that the node $c_1$ is not connected to the graph. It can be considered as a pilot symbol for the non-coherent setting.}
  \label{tanner_H}
\end{figure}

In this section, we focus again on the single user context, with $d$ reference symbols as per \textit{Case 1}.
Fig.~\ref{tanner_H} depicts the Tanner graph corresponding to \eqref{eq_code_non_coherent}.
White nodes denote information symbols from ($\mathbf{u}_s$), while red ones are codeword symbols from ($\mathbf{c})$. 
Purple ones are associated to noisy observed symbols from ($\mathbf{y})$.
Black squares denote constraint/function nodes.

In the factor graph from Fig. \ref{tanner_H}, 
the information variable nodes corresponding to symbols in mode $i$ are connected to $\prod_{n \neq i} T_{n} = T/T_i$ function nodes. 
Constraint nodes indexed by $\left(m_1, m_2, m_3\right)$ are connected to $d$ information variable nodes if $m_i \neq 1$ for all $i>1$.
Other constraint nodes exhibit smaller degrees due to the presence of reference symbols.
Generally, the number of constraint nodes with degree $r$ corresponds to the $r^{\text{th}}$ degree coefficient of the polynomial expansion of $\prod_{i=1}^{d}(1+(T_i-1)z)$.

Based on this graph representation, several approaches to apply message passing algorithms for approximate symbol MAP decoding are available. 
The first one consists in applying BP decoding based on discrete probability or log-probability messages \cite{SidharaPhD, SidharaTIT, Bariffi}, which is similar to the decoding of non-binary codes over finite fields. In the case of codes defined over $\mathbb{Z}_M$, due to the induced group structure, the check node update can be easily implemented using a FFT-BP approach as pointed out in \cite{SidharaPhD,GoupilTOC07}, where the fast Fourier transform (FFT) is the classical fast implementation of an $M$-point discrete Fourier transform (fast Hadamard transform for non binary codes over finite field). As finite field codes, the complexity increases with the field order. To overcome this problem,  \cite{Suresh_IZS2026_Riemannian_BP} has proposed a scalar BP approach based on the exchange of the complex von Mises moments (vM-BP), exploiting the structure of the PSK signaling that enables approximating the messages using von Mises probability density function.

\section{Performance Analysis}
This section investigates the performance of TBM-PSK under two frameworks: single-user AWGN and multi-user fading non-coherent multi-antenna scenarios. The objective is to characterize TBM-PSK as a code in terms of its reliability, rate efficiency, and scalability using state-of-the-art decoders. For the simulations, we consider \textit{Case 1}, having one reference symbol per tensor mode.

\subsection{Single-User Case}
We first analyze the performance of TBM-PSK in a single-user single antenna AWGN channel scenario. The received signal is modeled as $\mathbf{y} = \mathbf{s} + \mathbf{n}$,
where $\mathbf{y}, \mathbf{s}, \mathbf{n} \in \mathbb{C}^{T}$.
The block length is fixed to $T=3200$ channel uses.
SNR is calculated as $1/\sigma^2$, where $\sigma^2$ is the noise variance. 
\subsubsection{Rate Analysis} 
\begin{figure}[h]
    \centering
    \resizebox{0.45\textwidth}{!}{%
        \input{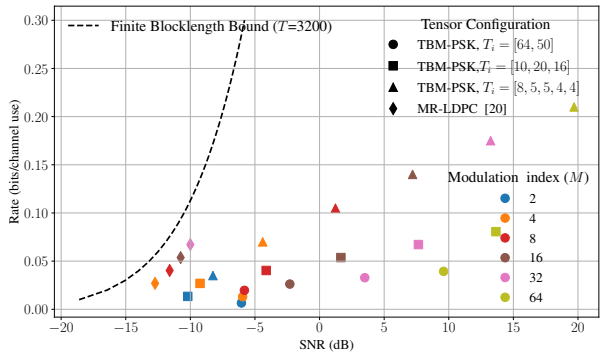}
    }
    \caption{Rate vs. minimum SNR required to achieve PER=$10^{-2}$.}
    \label{rate}
\end{figure}

Fig. \ref{rate} depicts the SNR required to get a packet error rate (PER) of $10^{-2}$ as a function of the code rate. 
For TBM-PSK, tensor configurations [64,50], [10,20,16] and [8,5,5,4,4] are considered with $M=2,4,8,16,32$ and 64.
Performance is evaluated using the vM-BP decoder from \cite{Suresh_IZS2026_Riemannian_BP}.

The considered TBM-PSK schemes are low-rate codes defined over the ring $\mathbb{Z}_M$. 
To give a fair benchmark to these types of codes, we have designed non-binary low-rate multiplicatively repeated LDPC (MR-LDPC) codes over $\mathbb{Z}_M$ that extend the design proposed by \cite{KasaiSNBLDPC} for non-binary LDPC codes over finite fields. 
This family of codes can be seen as a Raptor-like construction using a mother-based code that is extended by additional parity symbols to lower the rate. It has shown to be able to achieve good performance for very low rates. For our design, we have considered a simple mother code with rate $1/3$ over $\mathbb{Z}_M$, with regular variable nodes of degree $3$, and with dimension $K$. The number of added parity symbols is then selected to have $T$ coded symbols in order to have the same dimension and rate as the corresponding TBM-PSK scheme. Results for MR-LDPC are given for a maximum of $50$ decoding iterations using FFT-BP decoder over $\mathbb{Z}_M$ \cite{SidharaPhD,GoupilTOC07}.
Finally, the finite blocklength error bound with normal approximation derived in \cite[Sec. ~22.6]{polyanskiy2025information} for the AWGN channel is also depicted in Fig.~\ref{rate}.

\subsubsection{PER Analysis} 

\begin{figure}[h]
    \centering
    \resizebox{0.45\textwidth}{!}{%
        \input{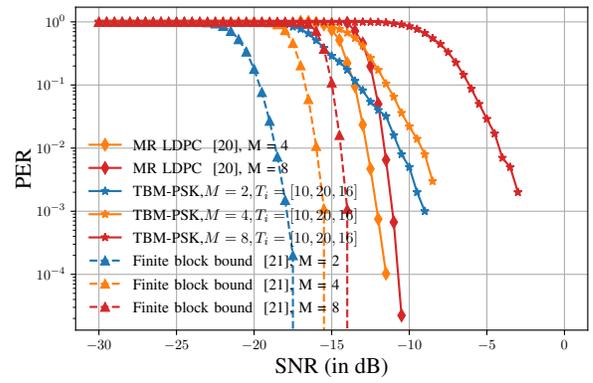}
    }
    \caption{PER comparison for $T = 3200$, AWGN channel.}
    \label{PER_vs_SNR}
\end{figure}

The PER is depicted versus SNR in Fig. \ref{PER_vs_SNR}, again for several configurations of TBM-PSK and MR-LDPC. The finite blocklength bound is also depicted. 
On this figure, curves with the same color correspond to the same coding rate.
Both figures \ref{rate} and \ref{PER_vs_SNR} show that TBM-PSK operates several dB away from the bound and from the MR-LDPC, indicating that they are very far from optimal as single-user codes for the AWGN channel. This however is not unexpected, given that they have been initially proposed in the context of random access over fading channels. Indeed, the true advantage of TBM emerges in the multi-user regime. 

\subsection{Multi-User Fading Channel Setup}
\begin{figure}[h]
    \centering
    \resizebox{0.4\textwidth}{!}{%
        \input{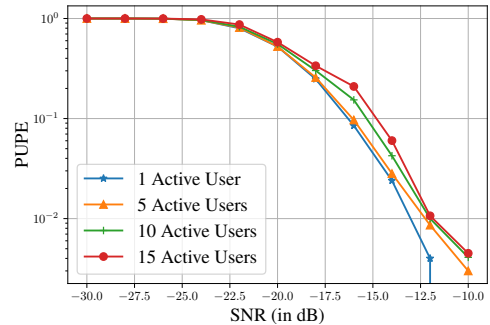}
    }
    \caption{PUPE vs. SNR, TBM-PSK $[10,20,16]$, $M=4$ over non-coherent quasi-static fading channel with $N_r=5$.}
    \label{PUPE_vs_Ka}
\end{figure}

We now focus on the multi-user non-coherent quasi-static fading scenario to examine the behavior of TBM-PSK under strong interference conditions, with multiple users transmitting simultaneously over the non-coherent, single-input multiple-output fading channel \eqref{TBM-s-h} with $N_r=5$ antennas.
Tensor configuration $T_i = [10,20,16]$ is considered with a 4-PSK.
The joint multi-user vM-BP decoder from \cite{Suresh_IZS2026_Riemannian_BP} is used here.
Fig. \ref{PUPE_vs_Ka} depicts the per-user probability of error (PUPE) \cite{Polyanskiy2017} as a function of SNR for $K_a =1,5,10$ and 15 active users.
The PUPE curve remains almost unchanged even though the number of active users $K_a$ increases up to 15, creating strong inter-user interference.
This indicates that the TBM-PSK code, together with the vM-BP decoding approach, is robust to strong interference levels, and can constitute an efficient basis for a parallel interference cancelation receiver.

\section{Conclusion}
In this paper, we presented a coding-theoretic interpretation of TBM with PSK modulation. We have shown that such codes can be decoded using the vM-BP approach, which is particularly beneficial when designing multi-user decoders performing parallel interference cancellation.
\bibliographystyle{IEEEtran}
\bibliography{bibliography.bib}

\end{document}